# Development stages of the "rope" human intestinal parasite

Alex A. Volinsky, Ph.D. [a*], Nikolai V. Gubarev, Ph.D. [b], Galina M. Orlovskaya, RN-C [c], Elena V. Marchenko, M.D., Ph.D. [a]

[a] Independent researcher
[b] Occupational Safety Ltd. (OOO "Bezopasnost Truda"), 32 ul. Koli Tomchaka, suite 14, St. Petersburg 196084, Russia
[c] Department of Surgery, St. Petersburg City Hospital No. 15, 4 Avangard St., St. Petersburg 198205, Russia

* Corresponding author. Phone: +1 813 974 5658, Fax: +1 813 974 3539
Email: alex.a.volinsky@gmail.com

**Abstract**

This paper describes the five development stages of the rope worm, which could be human parasite. Rope worms have been discovered as a result of cleansing enemas. Thousands or people have passed the rope worms from all over the World. Adult stages live in human gastro-intestinal tract and are anaerobic. They move inside the body by releasing gas bubbles utilizing jet propulsion. These worms look like a rope, and can be over a meter long. The development stages were identified based on their morphology. The fifth stage looks like a tough string of mucus about a meter long. The fourth stage looks similar, but the rope worm is shorter and has softer slimier body. The third stage looks like branched jellyfish. The second stage is viscous snot, or mucus with visible gas bubbles that act as suction cups. The first stage is slimier mucus with fewer bubbles, which can reside almost anywhere in the body. Rope worms have cellular structure, based on optical microscopy, DAPI staining and DNA analysis, however, the data collected is not sufficient to identify the specie. Removal methods are also mentioned in the paper.

**Disclaimer**

This research paper has not been peer reviewed. At the time of publication the authors have a hypothesis that the features described in this paper are of parasitic nature. Current DNA analysis results are inconclusive, however, only small percentage of the sequenced DNA has a match in GenBank. This paper was written for information purposes only, and is not intended to diagnose or treat any disease. If you are experiencing any symptoms, including those described in this paper, contact a licensed medical professional in your country.

**Introduction**

Human parasitic worms are classified as nematodes (roundworms), cestodes (tapeworms), trematodes (flukes) and monogeneans (Grove, 1990). It is estimated that every fourth human is hosting intestinal parasites (Watkins and Pollitt, 1997, World Development Report, 1993), meaning that even more people carry parasite intermediate stages. Humans can also carry intermediate stages of animal parasites, such as cat ascaris worms. Parasitic worms have different life cycles, sometimes using humans as permanent or temporary hosts. What if there is a parasite that does not have intermediate stages outside the human body, lives and dies with the human? Such specie, called rope worm, or *funis vermis* in Latin, has been recently discovered and described (Gubarev, 2009, Volinsky et. al. 2013). It does not fall under a single known parasite category. Based on its attributes, this pre-nematode may be older than other parasites. It could be a community of simple organisms forming a macroscopic organism, similar to biofilms and slime molds.

**Rope parasite adult 5$^{th}$ stage**

Figures 1 shows fully developed human rope worm passed with enemas from a 45 years old adult. These anaerobic parasites resemble human feces, and dry out outside the human body in air. They are called rope worms (*funis vermis* in Latin) because they look like twisted fibers of a rope (Figure 1). Rope worms color depends on the food a person eats, and varies from white to black. When a person is fasting, white worm leave the human body with enemas, so their original color may be white. Rope worms can be located almost anywhere in the human body, but prefer digestive tract, small and large intestines specifically. They twist like a corkscrew, increasing their cross-section, blocking the lumen of the intestine. This is also how rope worms squeeze the juice out of the fecal matter, and feed on it osmotically. To achieve this, the rope worm has multiple channels running along its length. The parasites emit gas bubbles inside these channels utilizing jet propulsion (Volinsky et. al. 2013). They are most active at night, between 1 and 6 am. High parasitic activity and toxins release can alter human attention and reaction.

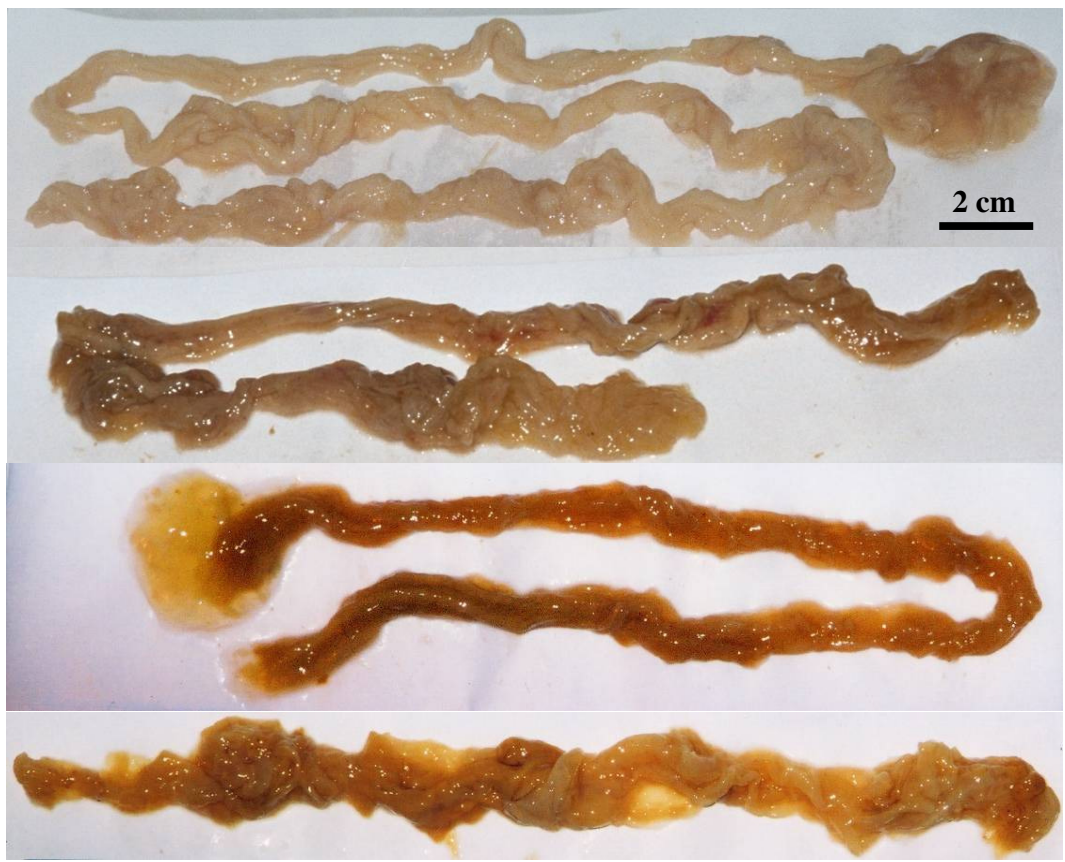

**Figure 1. Adult stages of the rope parasite (5$^{th}$ stage).**

Here are the reasons why the rope parasites can stay inside the human body without being carried out by peristaltic movements:
1) Rope parasites attach to the intestines with suction cups/heads;
2) Adult rope parasites reach over a meter in length, exceeding a typical length of the fecal contents;
3) Rope parasites move by emitting bubbles utilizing jet propulsion;
4) Rope parasites twist like a corkscrew and can completely block the lumen of the intestine;
5) Rope parasites form larger gas bubbles, which develop into suction cups.

The fifth adult stage can be driven away by enemas with eucalyptus decoction with several drops of eucalyptus oil, followed by the freshly squeezed lemon juice enema (Gubarev et. al. 2007).

**Rope parasite 4$^{th}$ stage**

The fourth stage looks similar to the 5$^{th}$ adult stage, but has softer slimier body (Figure 2). Both 5$^{th}$ and 4$^{th}$ stages can possibly feed on blood. They can emit bubbles to form future attachment heads, as seen in Figure 2. Similar to the 5$^{th}$ stage, the same eucalyptus/lemon juice enemas get rid of the 4$^{th}$ parasite stage (Gubarev et. al. 2007). Special care should be taken during the dehilminthation procedure, as open wounds are left on the inner side of the intestines, causing internal bleeding (see Figure 2). Bleeding can be stopped by using "dead" water made by water electrolysis.

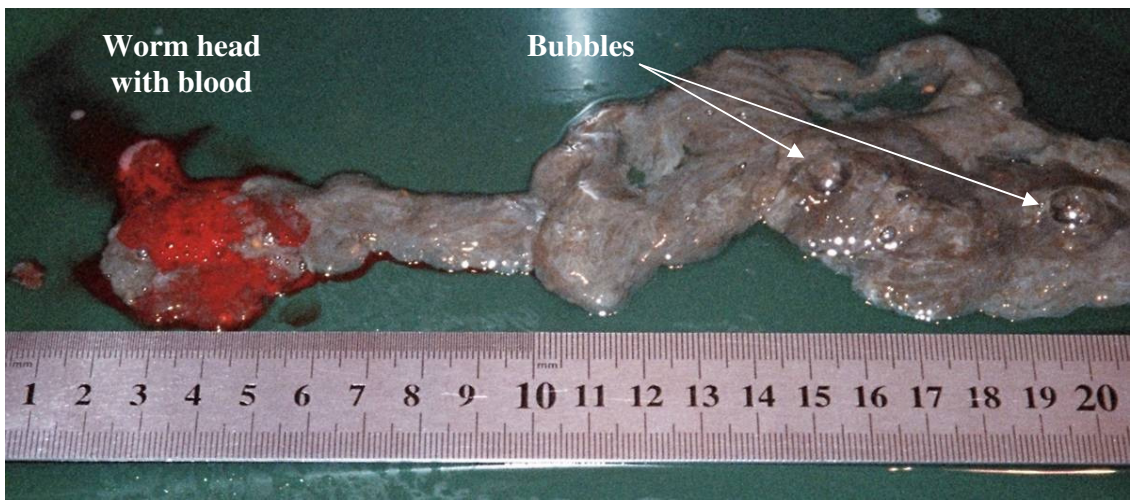

**Figure 2. Rope worm head covered with blood with multiple bubbles on the worm body.**

**Branched jellyfish 3$^{rd}$ stage**

The third stage looks like branched jellyfish, shown in Figure 3. Dehilminthation method includes enemas with baking soda (Gubarev et. al. 2006).

**Viscous mucus with bubbles 2$^{nd}$ stage**

The second stage resembles slimy viscous mucus, and emits bubbles, which are later used as attachment points (Volinsky et. al. 2013). This stage leaves the human body with salt milk enemas (Gubarev et. al. 2007).

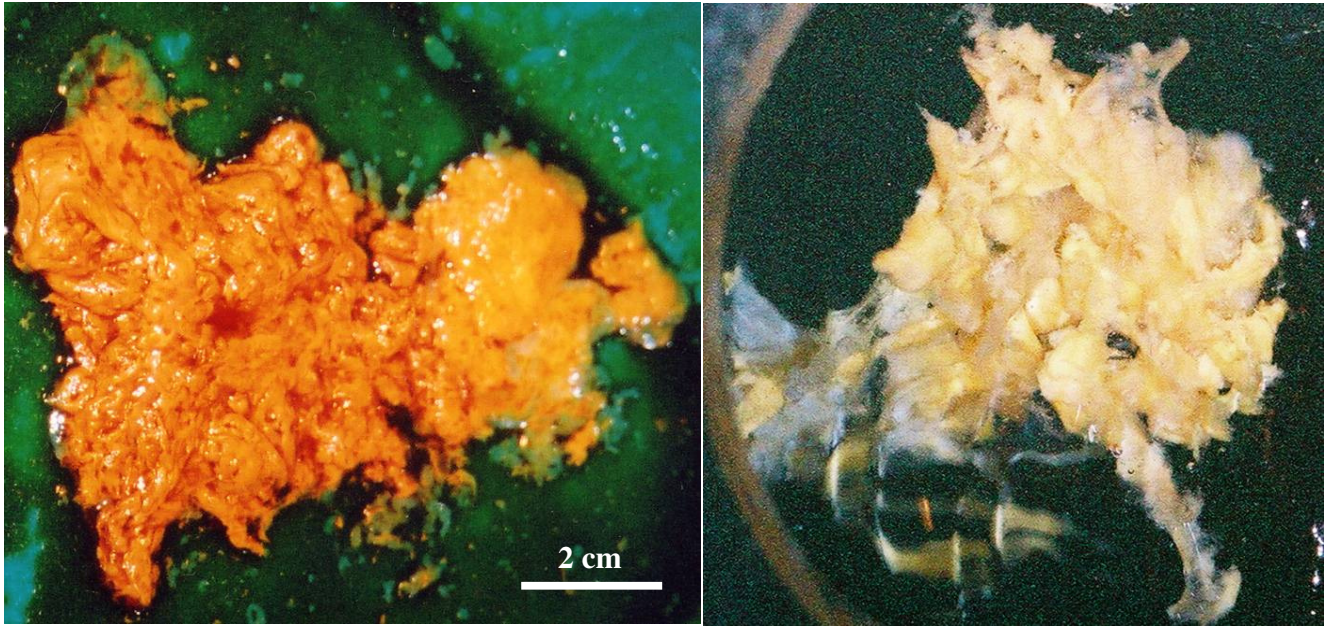

**Figure 3. Branched jellyfish 3$^{rd}$ stage of the rope parasite development.**

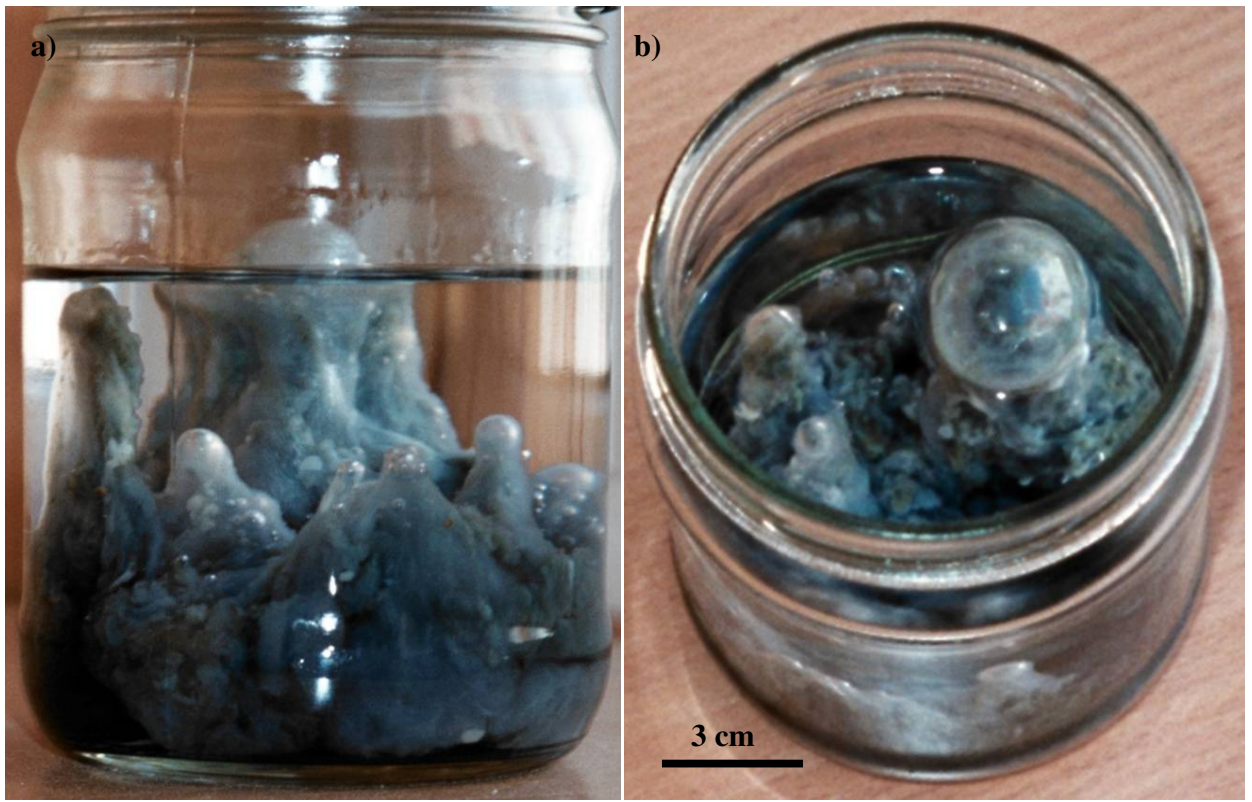

**Figure 4. Viscous mucus with bubbles 2$^{nd}$ development stage: a) side view; b) top view.**

**Viscous mucus 1st development stage**

The first stage of the rope parasites is mucus. It can be hosted almost anywhere in the human body. Similar to the second stage, salted milk enemas aid their release (Gubarev et. al. 2009).

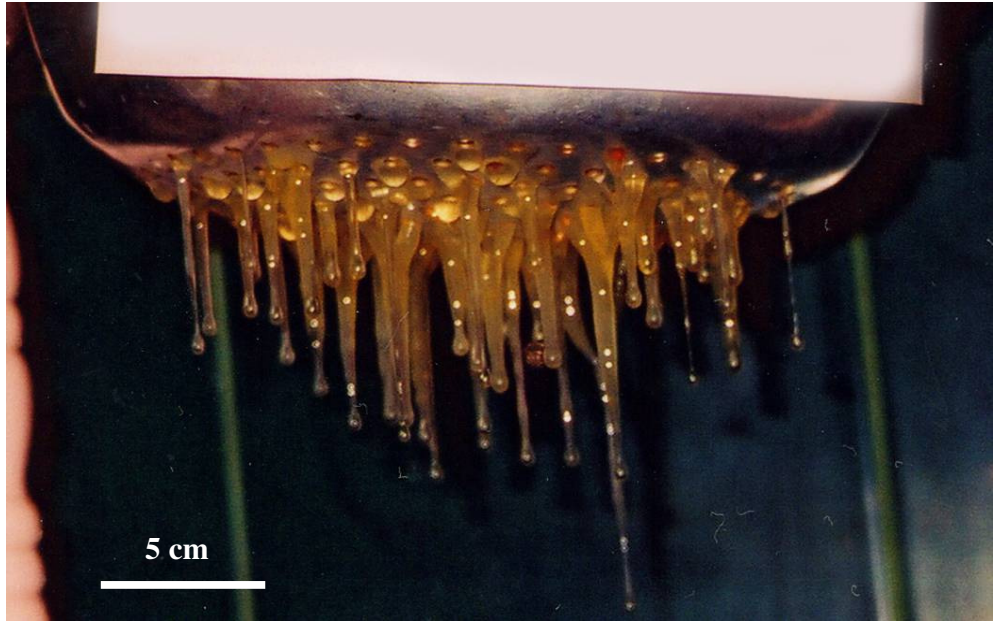

**Figure 5. Viscous mucus, the first development stage of the rope parasites, hanging from the colander holes.**

**Toxic slime and fecal stones**

Rope worms (stage 5) are also capable of producing toxic slime seen in Figure 6a. This happens when they are irritated by spicy food, heat or cold, etc. Adult rope worms also produce fecal stones, seen in Figure 6a and b. Fecal stones clearly have bright spots, which resemble sesame seeds, seen in Figure 6b. All fecal stoned from different people had these features. Fecal stones leave intestines with water enemas with small amounts of vinegar. Figure 6c shows an adult rope worm with the fecal stone attached to it. At this point it's not clear what the function of the fecal stones is, which could be reproductive or simply future food source storage.

**Discussion**

The first attempts to describe the structure of the adult rope worms, based on optical microcopy, revealed that they have multiple micochannels filled with gas bubbles (Volinsky et. al. 2013). The worm body is formed of the cells that resemble scales. The authors obtained scanning electron microcopy (SEM) images to better understand the rope parasite structure. Microchannels terminating on the worm surface were observed in SEM. Initially DNA analysis using primers was performed. COI gene sequences from the rope worms (stages 3 and 5) were obtained using Folmer primers. Stage 3 showed 99% match to human pseudogene, chromosomes 8 and 17. Stage 5 had 99% match to human mitochondria DNA. There is 82.6% match between the COI sequences obtained from the rope worm's stages 3 and 5. 18s gene sequences of stages 3 and 5 showed 99% human rRNA. There is 99.3% match between the two sequences. Illumina shotgun sequencing was also performed. Out of 15 Mbp sequenced less than 10% match bacterial and human DNA, while the rest currently has no match. Thus, the obtained DNA analysis results are inconclusive at this point.

Since the original publication in January 2013 over 200 people have contacted the authors, claiming that they are suffering from the rope worms and show pictures, similar to Figure 1. Others are Lyme disease patients and parents of autistic children, who pass these rope worms. Five people claim they have Morgellons disease, and also pass the rope worms. There are also videos of the rope worms moving in water, passed by an autistic child without any medications or procedures. Videos related the rope worms have been placed on the youtube.com channel (www.youtube.com/user/FunisVermis). There are also support groups on Facebook.com and several online forums.

The rope parasites have not been discovered previously for the following reasons:
1) Rope worms rarely come out as whole fully developed adult species;
2) Rope worms resemble human excrements;
3) Rope worms don't move outside the human body in air;
4) Rope worms are often mistaken for lining of the intestines.

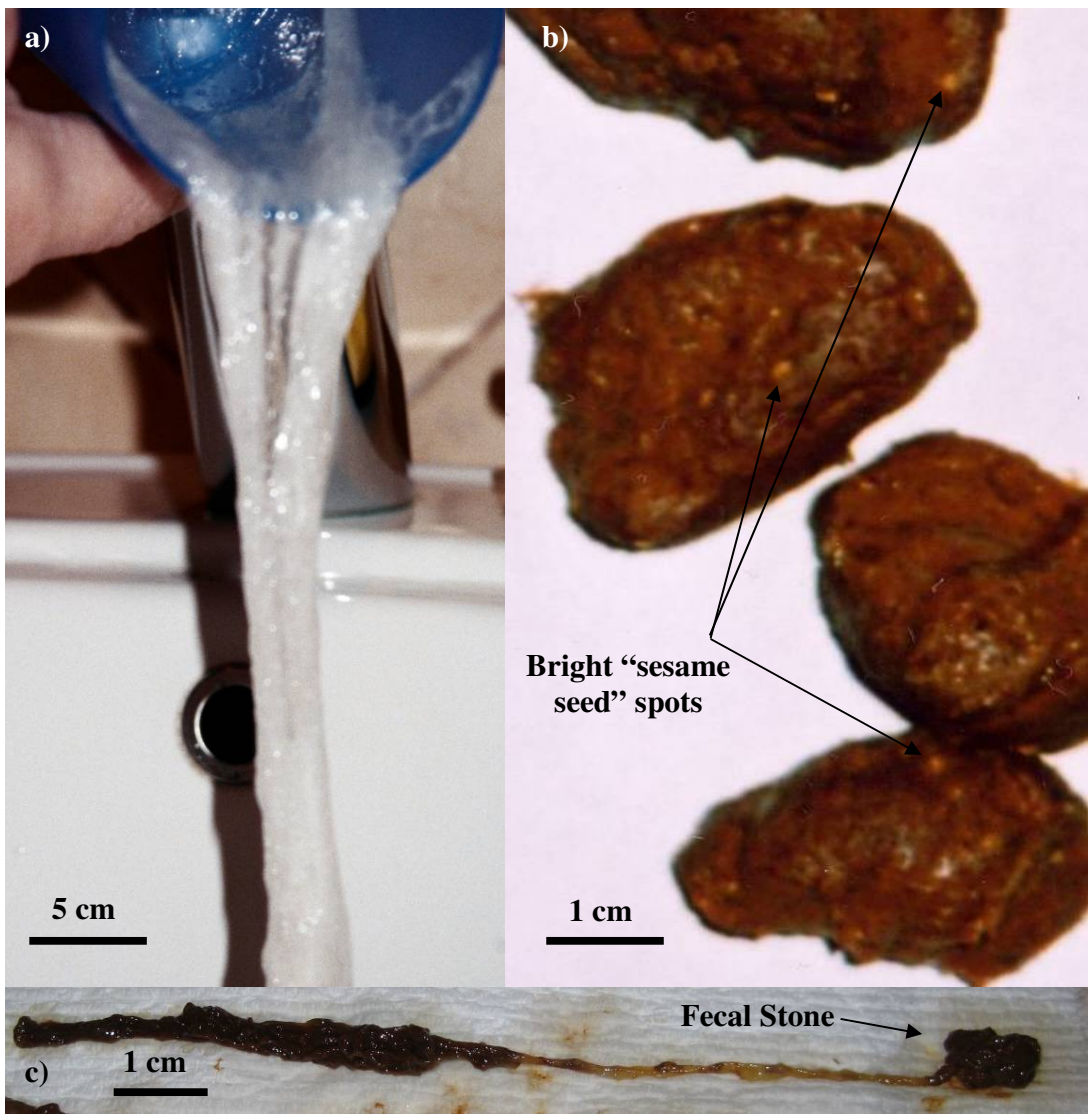

**Figure 6. a) Toxic slime produced by the rope worms; b) fecal stones produced by the rope worms; c) adult rope worm with a fecal stone attached.**

**Conclusions**

The five stages of the human anaerobic parasite, called the rope worm (*Funis Vermis* in Latin) have been described, based on the morphology. The currently known dehimnimthation methods include enemas with milk and salt, soda, eucalyptus, followed by the freshly squeezed lemon juice. Rope worms can possibly feed on human blood, thus special care should be taken upon dehilminthation to avoid internal bleeding. DNA analysis results are inconclusive at this point. Further research is required to identify what the rope worms are.

**Acknowledgements**

The authors acknowledge professionals support with microscopy and DNA data collection and analysis.